\documentclass[a4paper,10pt]{article}
\usepackage[utf8]{inputenc}
\usepackage[T1]{fontenc}
\usepackage[numbers,compress,merge]{natbib}
\usepackage[colorlinks=true]{hyperref}
\usepackage{graphicx}
\usepackage{amsmath,amssymb}
\usepackage{subfigure}
\usepackage{color,xcolor}
\usepackage[varg]{txfonts}
\usepackage[affil-it]{authblk}

\newcommand{\Ai}{A_{\omega l}^\text{in}}
\newcommand{\At}{A_{\omega l}^\text{tr}}
\newcommand{\Ar}{A_{\omega l}^\text{ref}}

\begin{document}
 
\title{Scalar scattering from black holes with tidal charge}

\author{Ednilton S. de Oliveira\footnote{ednilton@pq.cnpq.br}}
\affil{Faculdade de F\'isica, Universidade Federal do Par\'a,
	66075-110, Bel\'em, Par\'a, Brazil}

\date{\today}

\maketitle

\begin{abstract}
The cross sections of black holes with tidal charge predicted in the context of the Randall--Sundrum brane-world scenario are computed considering the massless scalar field. Results obtained for black holes with different tidal-charge intensities are compared in order to study how this charge modifies the black hole cross sections. Such results are also compared with the ones for Schwarzschild and extreme Reissner--Nordström black holes. The increase of the tidal-charge intensity makes the black hole absorb more and can also be measured by the narrowing of interference fringes of the differential scattering cross section. These results indicate that the effects of the tidal charge are very important in phenomena which take place near the black hole, but can be neglected in the far region. Analytical results are obtained in the high-frequency limit and are shown to excellently agree with the numeric results obtained via the partial-wave method. It is shown numerically that black holes with tidal charge obey the universality of the low-frequency absorption cross section of stationary black holes for the massless scalar field.
\end{abstract}


\section{Introduction}

Twenty years ago, Arkani-Hamed, Dimopoulos, and Dvali proposed that the fundamental scale of gravity could be so low as the weak scale if the spacetime had two or more extra compact dimensions~\cite{Arkani1998plb429_263}. In the same year, these authors, together with Antoniadis, showed that their proposal was based withing the context of string theory~\cite{Antoniadis1998plb436_257}. One year later, alternatives to this model were proposed by Randall and Sundrum considering that the spacetime has only one warped extra dimension~\cite{Randall1999prl83_3370,Randall1999prl83_4690}. Despite the differences, these scenarios are based on the fact that our 4-dimensional observed Universe is a subspace -- 3-brane, or simply brane -- of a higher-dimensional spacetime -- the bulk; Standard-Model fields are restricted to the 3-brane while gravity is free to penetrate the bulk. 

One of the most important consequences of these so called brane-world models is that microscopic black holes could be produced by particle collisions of a few TeVs at the LHC~\cite{Park2012ppnp67_617}.\footnote{See refs.~\cite{ATLAS2016jhep03_041,ATLAS2016plb754_302,CMS2017plb774_279,ATLAS2017prd96_052004,ATLAS2018epjc78_102} for some of the most recent constraints on quantum black hole production at the LHC.} Since these black holes would rapidly evaporate via Hawking radiation~\cite{Hawking1975cmp43_199}, some effort has been put on determining their greybody factors~\cite{Emparan2000prl85_499,Kanti2002prd66_024023,Kanti2003prd67_104019,Harris2003jhep10_014,Jung2004jhep09_005,Kanti2005prd71_104002,Ida2003prd67_064025,Ida2005prd71_124039,Ida2006prd73_124022,Harris2005plb633_106,Duffy2005jhep09_049,Casals2006jhep02_051,Casals2006jhep03_019,Creek2007prd75_084043,Kanti2014prd90_124077}, once their evaporation rate depend directly on these factors. However, these models have also important consequences on the astrophysical and cosmological levels~\cite{Majumdar2005ijmpd14_1095}, what instigated the appearing of some solutions which describe how black holes interact with particles restricted to the 3-brane (see refs. \cite{Park2012ppnp67_617,Bronnikov2003prd68_024025} and references therein).

Recently we have extended the investigation of black holes in brane-world models to their wave differential scattering cross sections, more specifically considering massless scalar plane waves impinging upon small static neutral and electrically charged black holes in the Arkani-Hamed--Dimopoulos--Dvali model~\cite{Marinho2016arxiv1612_05604,deOliveira2017cqg35_065007}. Other authors have previously considered some scattering properties of black holes on brane-world models, as the absorption cross section~\cite{Jung2004jhep09_005,Jung2005npb717_272,Toshmatov2016prd93_124017}, shadows~\cite{Schee2008ijmpd18_983,Amarilla2011pprd85_064019,Abdujabbarov2017prd96_084017,Eiroa2017epjc78_91} and deflections~\cite{Abdujabbarov2017prd96_084017,Kar2003grg35_1775} and also quasinormal modes~\cite{Toshmatov2016prd93_124017,Kanti2006prd73_044002,Kanti2006prd74_064008,Molina2016prd93_124068}.

In this work we present the study of the cross sections of a black hole with tidal charge~\cite{Dadhich2000plb487_1} predicted as a solution considering the Randall--Sundrum scenario. This system is described by the following metric on the 3-brane:\footnote{Here we work with $c = G = 1$.}
\begin{equation}
 ds^2 = f(r)dt^2-f(r)^{-1}dr^2 - r^2(d\theta^2+\sin^2\theta d\phi^2),
 \label{metric}
\end{equation}
where
\begin{equation}
 f(r)= 1 - \frac{2M}{r} + \frac{\beta}{r^2},
\end{equation}
with $M$ being the black hole mass and $\beta = qM^2$. If $0 < q \le 1$, then the metric~\eqref{metric} coincides with the one of Reissner--Nordström black holes~\cite{Chandra1983}; $q = 0$ describes Schwarzschild black holes; negative values for $q$ are not allowed in the context of General Relativity, but are permitted in the context of the Randall--Sundrum brane-world scenario where $\beta$ represents the effect of the 5th dimension on the 3-brane~\cite{Dadhich2000plb487_1} and is related to as ``tidal charge''. The event horizon of such systems is given by
\begin{equation}
r_h = M\left(1 + \sqrt{1-q}\right).
\label{rh}
\end{equation}
We see that the tidal charge acts in the same sense of the black hole mass, increasing the size of the event horizon, while the electric charge on the Reissner--Nordström solution acts in the opposite sense, once $r_h$ tends to decrease with the increase of $q$. Other consequences of the tidal charge have been studied in detail. For example, in ref.~\cite{Toshmatov2016prd93_124017} it has been shown that the tidal charge tends to decrease the oscillation frequency while increase the damping rate of scalar, electromagnetic, and gravitational quasinormal modes. The authors also showed that the absorption cross section and the emission rate of such black holes increase with the increase of the tidal charge. In refs.~\cite{Schee2008ijmpd18_983,Amarilla2011pprd85_064019,Abdujabbarov2017prd96_084017,Eiroa2017epjc78_91} the influence of the tidal charge in the shadow cast by a black hole was studied with the conclusion that the shadow increases if the tidal-charge intensity increases.

Here our interest lies in the roles the tidal charge plays in the cross sections of black holes described by eq.~\eqref{metric}. We use the massless scalar field to model a plane wave impinging upon the black hole and then compute its absorption and differential scattering cross sections numerically using the partial-wave method. We compare results for black holes with different tidal charges and also for Schwarzschild and extreme Reissner--Nordström black holes in order to better understand the effects of the tidal charge on the scattering properties of these black holes.

The present work is organized as follows: in sec.~\ref{sec:hf} we study the high-frequency limits of the absorption and scattering cross sections; in sec.~\ref{sec:wave_scatt} we describe the behavior of the massless scalar field in the considered spacetime in terms of partial waves, presenting asymptotic solutions to the Klein-Gordon equation, as well as the general expressions for the cross sections; section~\ref{sec:results} shows a selection of cross sections obtained numerically and also their comparisons with the respective analytical results; our final remarks are presented in sec.~\ref{sec:remarks}.

\section{Classical and semi-classical scattering\label{sec:hf}}

\subsection{Geodesic limit}

The spacetime described in eq.~\eqref{metric} is spherically symmetric. Therefore, the particle motions have two conserved quantities given by
\begin{equation}
E = f\dot{t}
\label{E}
\end{equation}
and
\begin{equation}
L = r^2\dot{\phi},
\label{L}
\end{equation}
which are related respectively to the energy and the angular momentum of the particle. The dot means differentiation with respect to an affine parameter. Such constants can be used to describe the impact parameter of scattered particles, $b = L/E$. The deflection of null geodesics can be given in terms of the following equation:
\begin{equation}
 \left(\frac{du}{d\phi}\right)^2 = \frac{1}{b^2} - u^2f(1/u) = h_b(u),
 \label{def_ang1}
\end{equation}
where $u\equiv 1/r$. We can obtain a second-order equation with the differentiation of the above equation:
\begin{equation}
 \frac{d^2u}{d\phi^2} + u = 3Mu^2-2\beta u^3.
\label{def_ang2}
\end{equation}
By making $d^2u/d\phi^2=0$ we obtain the critical-orbit radius
\begin{equation}
r_c = -\frac{4\beta}{-3M+\sqrt{9M^2-8\beta}}.
\label{rc}
\end{equation}
From $h_b(1/r_c) = 0$, we find the critical impact parameter $b_c = r_c/f(r_c)^{1/2}$ which is the radius of the capture cross section of the black hole, $\sigma_\text{abs}^{\text{cl}} = \pi b_c^2$.

The deflection angle can be obtained by integrating eq.~\eqref{def_ang1} from $u = 0$ to $u=1/r_0$, being $r_0$ the radius of the returning point. For $\beta < 0$ this leads to
\begin{equation}
 \Theta(b) = \frac{4}{\sqrt{-\beta(u_1-u_4)(u_2-u_3)}} [K(k)-F(z,k)],
 \label{Th}
\end{equation}
where $K(k)$ and $F(z,k)$ are the complete and incomplete elliptic integrals of the first kind~\cite{Abramowitz_etal1964} and their arguments are
$$
z = \sqrt{\frac{u_3(u_1-u_4)}{u_4(u_1-u_3)}}
\qquad \text{ and } \qquad 
k = \sqrt{\frac{(u_1-u_3)(u_2-u_4)}{(u_2-u_3)(u_1-u_4)}}.
$$
Here, $u_i$ ($i = 1\ldots 4$) are the roots of $h_b(u)$ for $b \ge b_c $ with $u_2 \ge u_1 > 0$ and $u_4 < u_3 < 0$; $u_1 = 1/r_0$. Although expression~\eqref{Th} is the same as in the Reissner--Nordström case, now $h_b(u)$ admits two negative roots for $b > b_c$, while in the former case it has only one~\cite{Crispino2009prd79_064022}.

The classical differential scattering cross section is given by
\begin{equation}
 \frac{d\sigma_\text{el}^\text{(cl)}}{d\Omega} = \sum_b \frac{b}{\sin\theta} \left| \frac{db}{d\theta} \right|,
 \label{geo_scs}
\end{equation}
where the sum takes the fact that particles can circulate the black hole multiple times so that the same scattering angle $\theta = |2n\pi - \Theta|$ ($n = 0,1,2\ldots$) can be observed for particles with different impact parameters, e.g. different values of $n$.

\begin{figure}[htpb!]
\centering
 \includegraphics[width=0.49\textwidth]{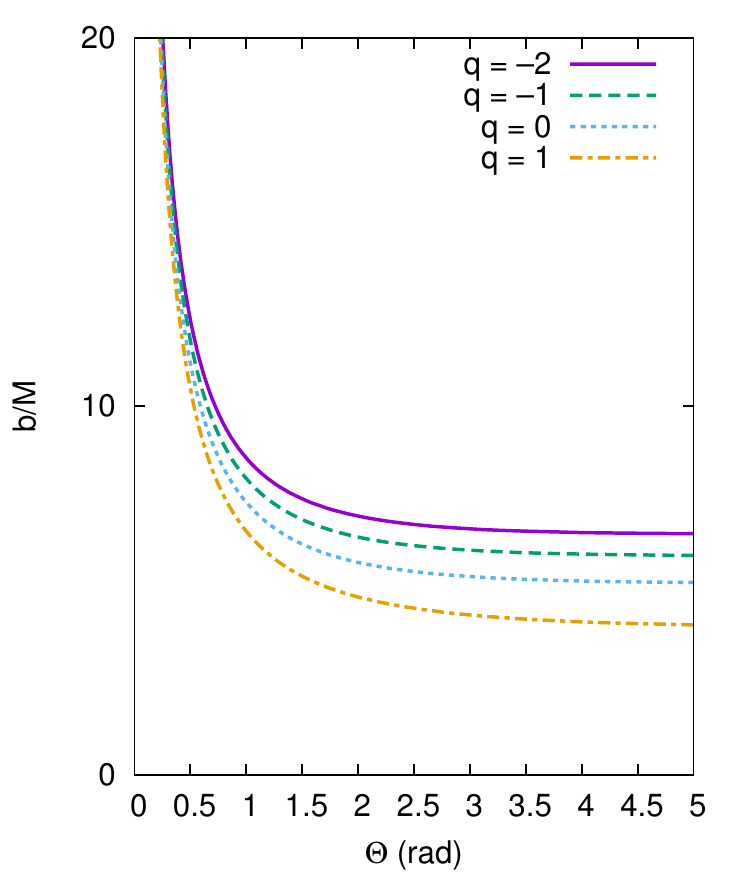}
 \includegraphics[width=0.49\textwidth]{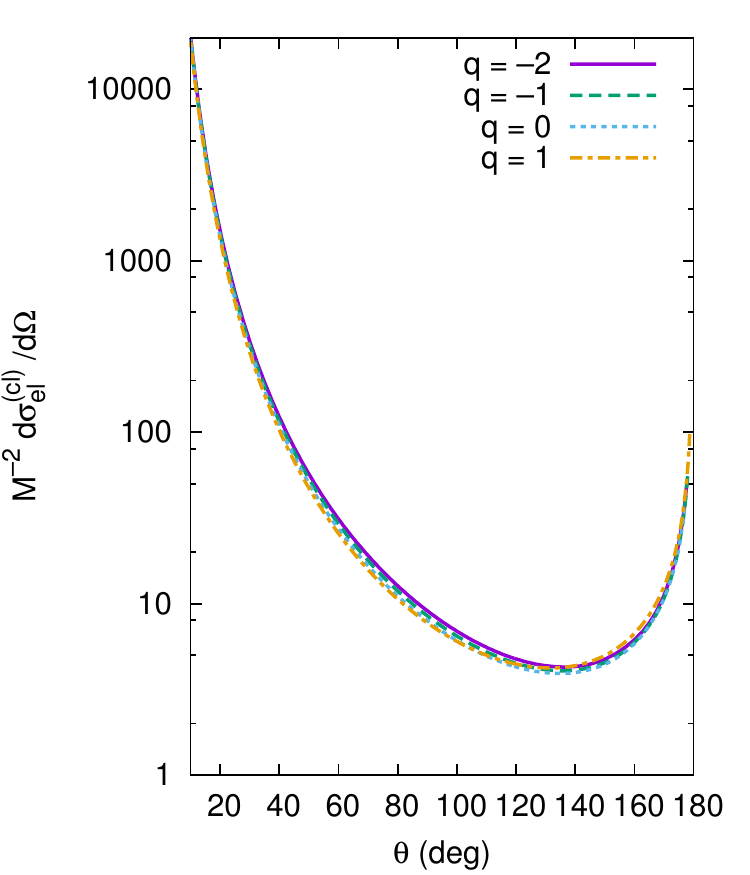}
 \caption{Comparison of the deflection angle (left) and the null-geodesic differential scattering cross section (right) for black holes with $q = -2,-1,0,1$.}
 \label{fig:geo_scatt}
\end{figure}

Figure~\ref{fig:geo_scatt} shows the deflection angle and the differential scattering cross section for massless particles considering $q = -2,-1,0,1$. The differential scattering cross section has been computed considering until the second largest term of the sum in eq.~\eqref{geo_scs} dropping its contribution only when it was smaller than $\sim 0.1 \%$; the contributions of further terms are even smaller. We see that particles can approach closer black holes with higher $q$ and that the charge exerts lower influence in particles scattered with high impact parameters. As consequence, it becomes hard to make distinction between the differential scattering cross sections of black holes with different charges when considering particles scattered in small-angle directions ($\theta \lesssim 20^\circ$). As we see in sec.~\ref{sec:results}, these results agree very well with the wave differential scattering cross section in the same limit they tend to each other.

\subsection{Glory approximation}

Near the backward direction ($\theta = 180^\circ$), the scalar scattering cross section can be described analytically by the \emph{glory} approximation~\cite{DeWitt-Morette1984prd29_1663,Futterman_etal1988}:
\begin{equation}
 \frac{d\sigma_\text{el}^{(\text{gl})}}{d\Omega} = 2\pi \omega b_g^2 \left| \frac{db}{d\theta} \right|_{\theta = \pi} J_0^2(b_g\, \omega \sin\theta),
\label{gl_scs}
\end{equation}
where $J_n(\cdot)$ is the Bessel function of the first kind~\cite{Gradshteyn_etal2000}, and $b_g$ is the impact parameter of back-scattered rays. Equation~\eqref{gl_scs} is a semi-classical formula, as we can infer from the fact that it involves both ray ($b$) and wave ($\omega$) properties. Therefore, it is expected to be valid for $M\omega \gg 1$. Despite this, the glory approximation presents very good agreement when compared with the numeric results for intermediate values of frequency, for instance $M\omega = 5.0$, as we will see below.

\begin{figure}[!htb]
 \centering
 \includegraphics[width=0.49\textwidth]{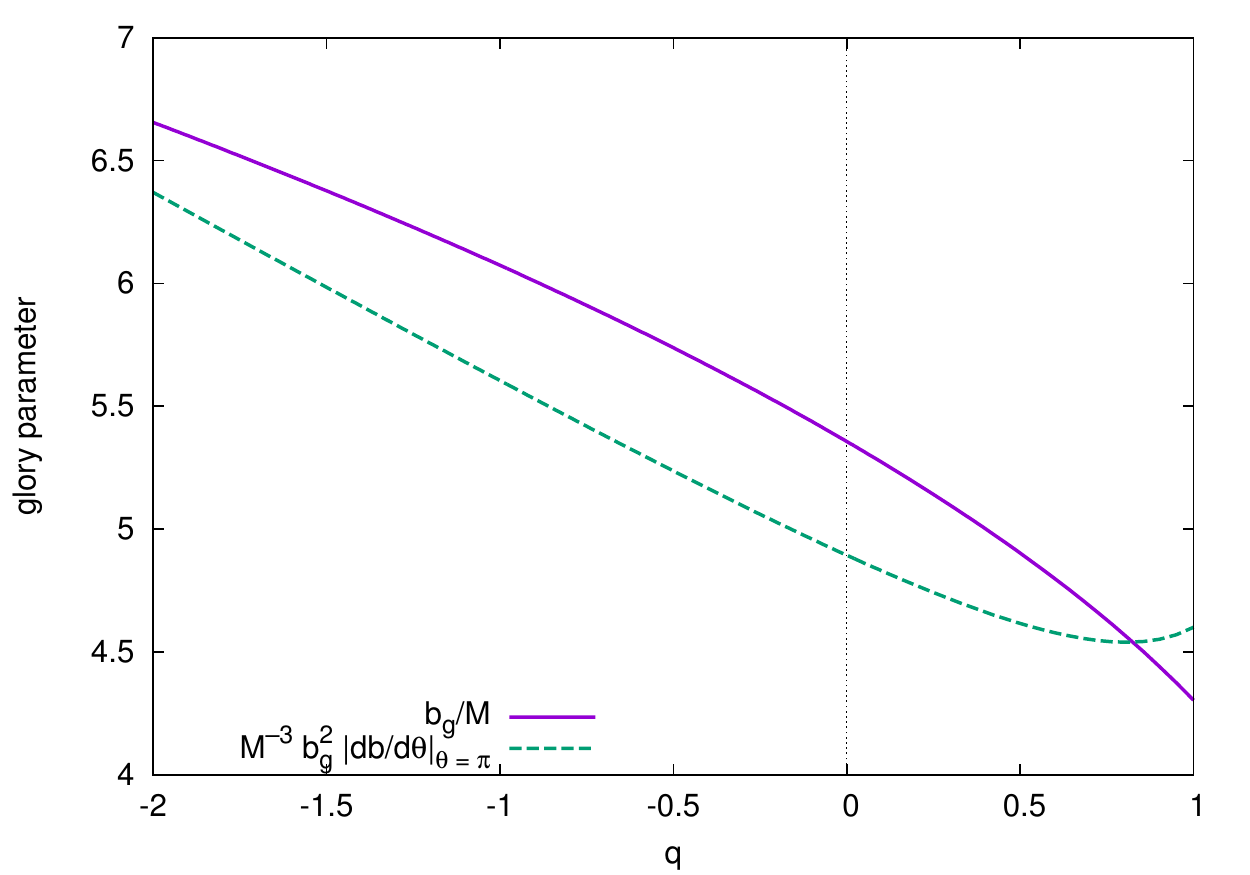}
 \caption{The classical parameters which govern the glory ring widths and intensities as functions of $q$.}
 \label{fig:gl_params}
\end{figure}

In fig.~\ref{fig:gl_params} we plot the classical parameters which define the intensity, $b_g^2 |db/d\theta|_{\theta = \pi}$, and the widths, $b_g$, of the glory rings. As we can see, the glory intensity increases with the increase of the tidal-charge intensity, i.e., as $|q|$ becomes higher on the $q < 0$ region. This is not observed in the case of Reissner--Nordström black holes, where the glory intensity decreases with the increase of black hole electric charge up to $q \approx 0.8$ when the intensity starts increasing~\cite{Crispino2009prd79_064022}.\footnote{The dashed line in the region $q \ge 0$ of fig.~\ref{fig:gl_params} is equivalent to the solid line of fig.~10 of ref.~\cite{Crispino2009prd79_064022}. However, it is important to note that $q$ in ref.~\cite{Crispino2009prd79_064022} is proportional to the black hole electric charge intensity, $|Q|$, while here $q$ is defined as being proportional to $Q^2$. Therefore, although both lines represent the same quantity, apart from a multiplicative factor, their shapes look different.} Formula~\eqref{gl_scs} predicts that the interference fringe widths vary inversely to $b_g$. Therefore, from fig.~\ref{fig:gl_params} we conclude that the fringe widths must decrease with the increase of the tidal-charge intensity. In the Reissner--Nordström case, the increase of the black hole charge intensity results in a increase of the interference fringe widths~\cite{Crispino2009prd79_064022}. As we see in sec.~\ref{sec:results}, this last prediction based on the glory approximation is confirmed by the numeric results.

In sec.~\ref{sec:results} we compare the glory approximation with the numeric results obtained via the partial-wave method. This comparison is important because it can help us to estimate the accuracy of our results or, in some cases, the lack of agreement between the glory approximation and the partial-wave results may indicate the occurrence of extraordinary phenomena in the scattering process. This is the case of the electromagnetic and gravitational scatterings from Reissner--Nordström black holes, where both helicity-reversing process and interconversion of spin 1 and 2 waves take place~\cite{Crispino2014prd90_064027,Crispino2015prd92_084056}.


\section{Wave scattering\label{sec:wave_scatt}}

Here we consider the case of the massless scalar field model to describe the scattered wave. This field is governed by the Klein--Gordon equation, which reads:
\begin{equation}
 \frac{1}{\sqrt{-g}} \partial_{\mu} \left ( \sqrt{-g} g^{\mu\nu} \partial_\nu \Phi \right)  = 0.
 \label{KG}
\end{equation}
The metric in this equation is implicitly given in eq.~\eqref{metric}, with $g$ denoting its determinant. Spherical symmetry of the spacetime allows us to define stationary modes proportional to the scalar spherical harmonics as $\Phi_{\omega lm} = [\psi_{\omega l}(r)/r] Y_l^m(\theta,\phi) e^{-i\omega t}$. Once angular and temporal parts of the solution are known, we have to focus on solving its radial part, which can been shown to be:
\begin{equation}
 f \frac{d}{dr}\left(f \frac{d\psi_{\omega l}}{dr} \right) + \left[\omega^2 - V_l(r) \right ] \psi_{\omega l} = 0,
 \label{radial_eq}
\end{equation}
where the effective potential is given by
\begin{equation}
 V_l(r) = f\left[\frac{1}{r}\frac{df}{dr} + \frac{l(l+1)}{r^2} \right].
 \label{V}
\end{equation}
The radial equation~\eqref{radial_eq} can be put in a Schrödinger-like form if we define the tortoise coordinate $d/dr_* = f\, d/dr$:
\begin{equation}
 \frac{d^2 \psi_{\omega l}}{dr^2_*} + \left[\omega^2 - V_l(r_*) \right ] \psi_{\omega l} = 0.
 \label{radial_eq-tor}
\end{equation}

\begin{figure}[!htb]
\centering
\includegraphics[width=0.49\textwidth]{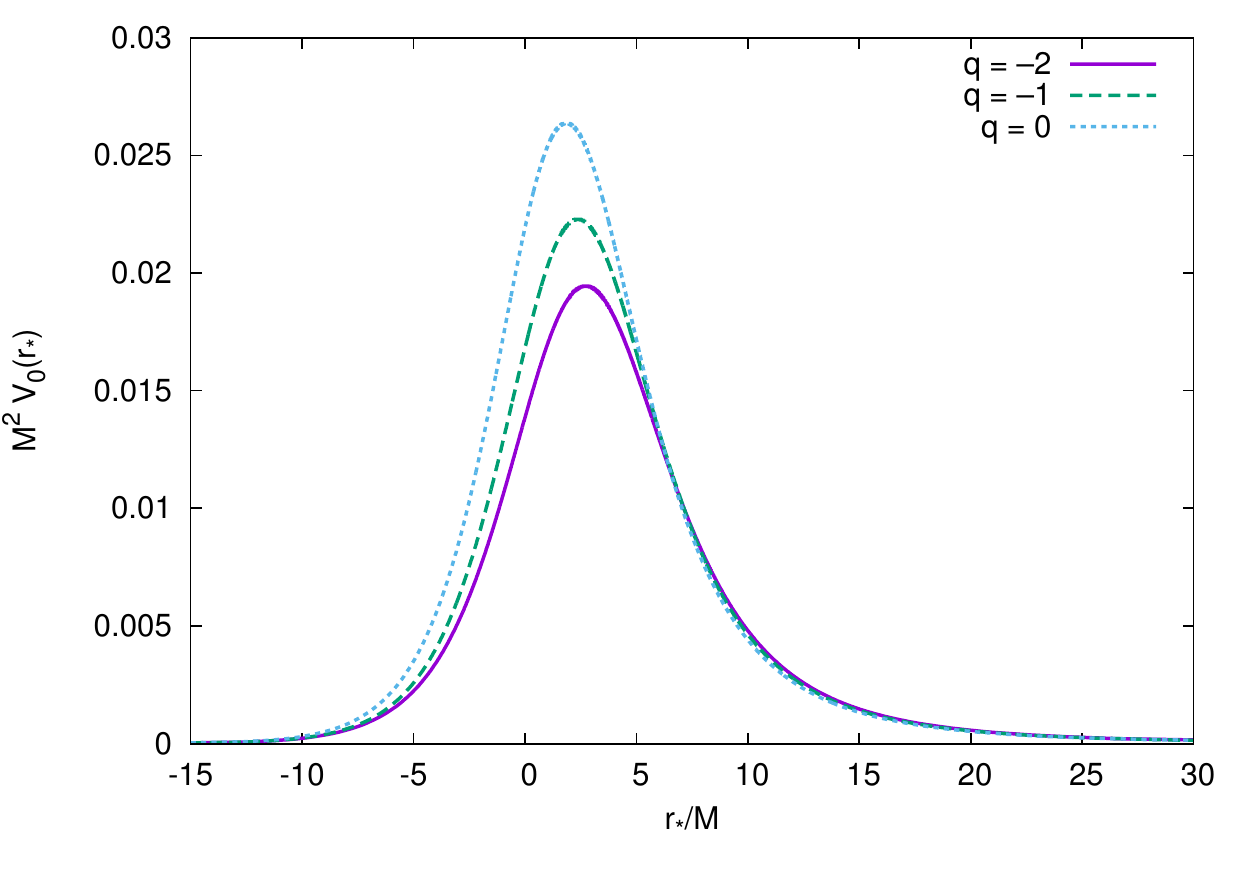}
\includegraphics[width=0.49\textwidth]{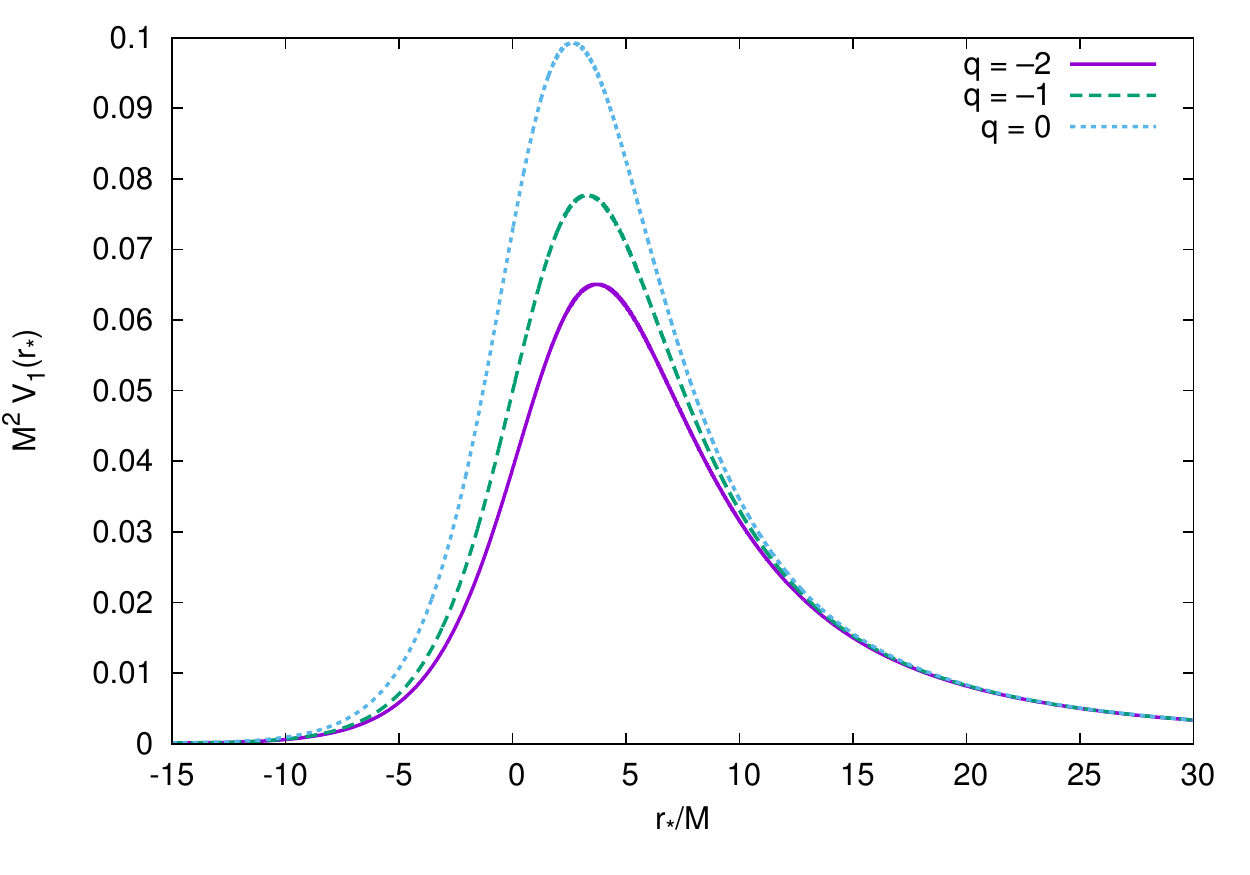}
\caption{Effective potential with $l = 0$ (left) and $l = 1$ (right) in terms of the tortoise coordinate for black holes with $q = 0, -1, -2$. The effective potential barrier increases with the increase of $l$, but its shape remains similar to the case $l = 1$ (right panel).}
\label{fig:V}
\end{figure}

Some plots of the effective potential~\eqref{V} are shown in fig.~\ref{fig:V}. There we can see that the effective potential vanishes in the limits $r_* \to \pm \infty$ independently of the value of $q$ or $l$. The main consequence of changing $q$ is observed in the maximum value of the effective potential, which decreases with the increase of the tidal charge intensity, $|q|$. Therefore, we can already predict that partial waves are more absorbed by black holes which have a more intense tidal charge. This agrees with what has been previously observed in the behavior of the absorption cross section in ref.~\cite{Toshmatov2016prd93_124017}, the black hole shadows in refs.~\cite{Schee2008ijmpd18_983,Amarilla2011pprd85_064019,Abdujabbarov2017prd96_084017,Eiroa2017epjc78_91}, and also with the high-frequency analysis presented in sec.~\ref{sec:hf} where we observed that massless particles can come closer to the black hole without being absorbed as higher is the value of $q$ (see fig.~\ref{fig:geo_scatt}, left panel).

The asymptotic analysis of $V_l(r_*)$ allows us to describe the behavior of $\psi_{\omega l}$ in regions near the horizon and far from the black hole. Such behavior is necessary to provide the mathematical expressions of the cross sections. For $r_* \to -\infty$, $V_l(r_*) \to 0$ and therefore, for the scattering problem
\begin{equation}
 \psi_{\omega l} \approx \At \,  e^{-i\omega r_*}.
 \label{psi_hor}
\end{equation}
For $r_* \to \infty$, $V_l(r_*) \to 0$ and therefore
\begin{equation}
 \psi_{\omega l} \approx \Ai \, e^{-i\omega r_*} + \Ar \, e^{i\omega r_*}.
 \label{psi_inf}
\end{equation}
In the region $r \gg r_h$, a more precise form of the radial function can be obtained by considering that $V_l(r_*) \approx l(l+1)/r_*^2$ ($r_* \sim r$). In this case, the radial solution can be expressed as:
\begin{equation}
\psi_{\omega l}(r_*) \approx \omega r_*\left[(-i)^{l+1}\Ai \, h_l^{(2)}(\omega r_*) + i^{l+1} \Ar \, h_l^{(1)}(\omega r_*)\right],
\label{psi_hankel}
\end{equation}
where $h_l^{(1,2)}(\cdot)$ are the spherical Hankel functions of the first and second kind~\cite{Abramowitz_etal1964}, respectively. Once $h_l^{(1)} (x) \approx (-i)^{l+1} e^{ix}/x$ and $h_l^{(2)}(x) \approx i^{l+1} e^{-ix}/x$ in the region $x \gg l(l+1)/2$, we recover~\eqref{psi_inf} from~\eqref{psi_hankel} at infinity, as expected.

We can define the reflection and transmission coefficients in terms of $\Ai$, $\Ar$, and $\At$ respectively as
\begin{equation}
 \mathcal{R}_{\omega l} \equiv \left|\frac{\Ar}{\Ai}\right|^2,
 \label{R}
\end{equation}
and
\begin{equation}
 \mathcal{T}_{\omega l} \equiv \left|\frac{\At}{\Ai}\right|^2.
 \label{T}
\end{equation}
Flux conservation implies in $\mathcal{R}_{\omega l} + \mathcal{T}_{\omega l} = 1$. These coefficients are also necessary in the description of the absorption cross section, which for massless scalar monochromatic plane waves scattered in spherically symmetric spacetimes can be shown to be~\cite{Jung2005npb717_272,Sanchez1976jmp17_688,Sanchez1976prd16_937,Sanchez1978prd18_1030,Crispino2007prd76_107502,Macedo2014prd90_064001}:
\begin{equation}
 \sigma_\text{abs} = \sum\limits_{l = 0}^{\infty} \sigma_\text{abs}^{(l)},
 \label{tacs}
\end{equation}
where
\begin{equation}
 \sigma_\text{abs}^{(l)} = \frac{\pi}{\omega^2} (2l+1)\mathcal{T}_{\omega l}
 \label{pacs}
\end{equation}
is the absorption cross section of each partial wave, usually referred to as partial absorption cross section.

We can also define the phase shifts from the coefficients given in eq.~\eqref{psi_inf}:
\begin{equation}
 e^{2i\delta_l(\omega)} = (-1)^{l+1} \frac{\Ar}{\Ai}.
 \label{ps}
\end{equation}
The differential scattering cross section for massless scalar monochromatic plane waves in spherically symmetric spacetimes can be given as~\cite{Crispino2009prd79_064022,Sanchez1978prd18_1798,Dolan2009prd79_064014,Macedo2015prd91_024012}:
\begin{equation}
 \frac{d\sigma_\text{el}}{d\Omega} = |f_\omega (\theta)|^2,
 \label{dscs}
\end{equation}
where $f_\omega(\theta)$ is the scattering amplitude which in terms of partial waves is:
\begin{equation}
 f_\omega(\theta) = \frac{1}{2i\omega} \sum\limits_{l=0}^{\infty} (2l+1) \left[ e^{2i\delta_l(\omega)} - 1\right] P_l (\cos\theta),
 \label{amp}
\end{equation}
with $P_l(\cdot)$ being the Legendre polynomials.

The sum of the total absorption cross section $\sigma_\text{abs}$ with the scattering cross section $\sigma_\text{el}$ defines the total cross section, $\sigma_\text{tot}$. The total cross section is known to diverge if the wave is scattered by potentials which asymptotically fall off as the Coulomb potential, $V \sim 1/r$. This is the case of the spacetime studied here, since the main term in the metric at infinity comes from the Schwarzschild term, $-2M/r$. However, it is possible to obtain a finite total cross section for small black holes on the brane considering the ADD model if the bulk has 6 or more dimensions~\cite{Marinho2016arxiv1612_05604}.

\section{Numeric results\label{sec:results}}

Here we present numeric results for the absorption~\eqref{tacs} and differential scattering cross sections~\eqref{dscs}. These results are obtained by matching numeric solutions of the radial eq.~\eqref{radial_eq} with the corresponding asymptotic solutions, eq.~\eqref{psi_inf}. In order to improve precision, we may use the solution in terms of spherical Hankel functions, eq.~\eqref{psi_hankel}, or improve the asymptotic solutions~\eqref{psi_hor} and \eqref{psi_inf} with a power series expansion (\emph{cf.}, for example, eqs.~(15) and (16) of ref.~\cite{Macedo2015prd91_024012}); here we use the first approach. Once the transmission coefficients are found, the computation of the absorption cross sections is straightforward. The situation is not so direct in the case of the differential scattering cross section. The scattering amplitude sum~\eqref{amp} is known to be divergent in the forward direction and poorly convergent in other cases. Therefore, we apply a convergence method introduced in ref.~\cite{Yennie1954pr85_500} in order to obtain precise results for the differential scattering cross section computing a relatively small number of phase shifts.

\begin{figure}[!htpb]
\centering
 \includegraphics[width=0.49\textwidth]{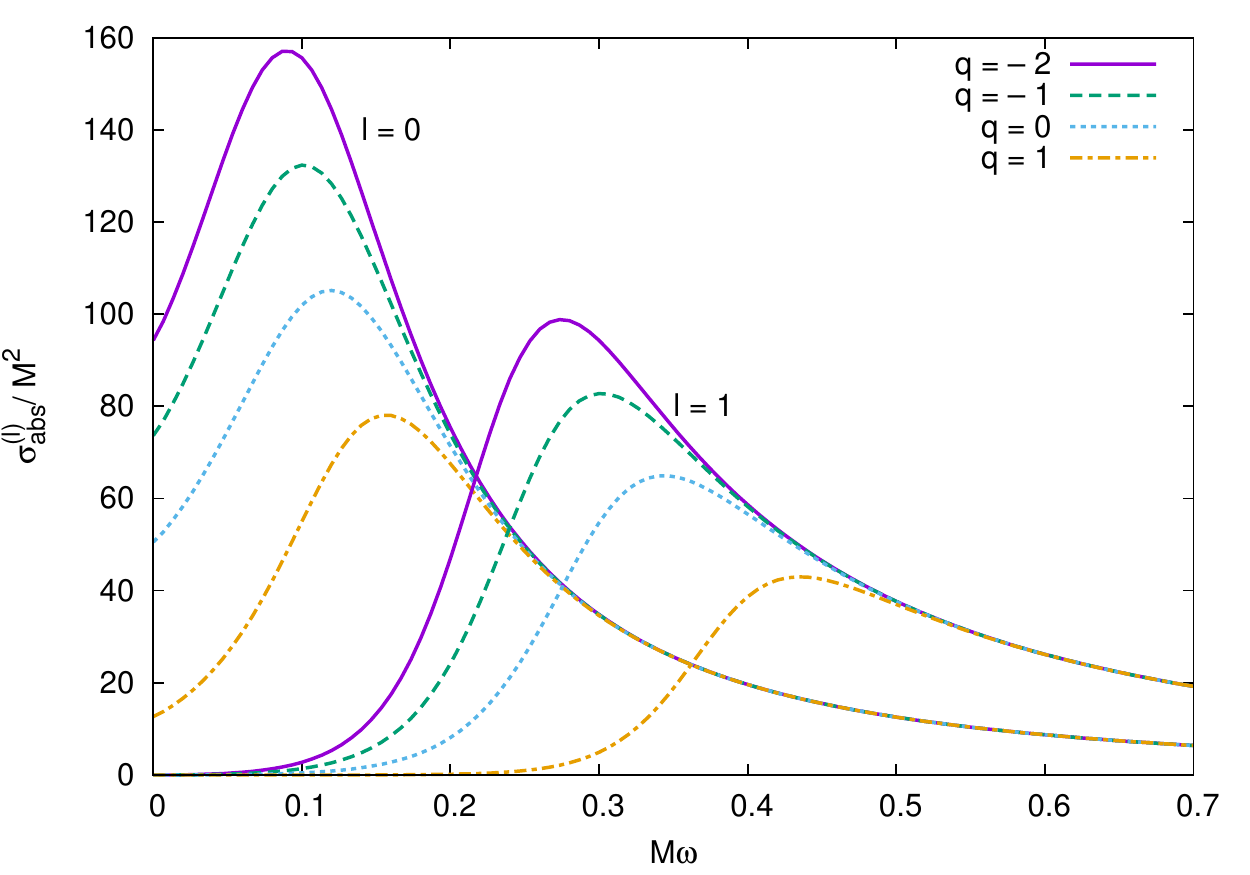}
 \includegraphics[width=0.49\textwidth]{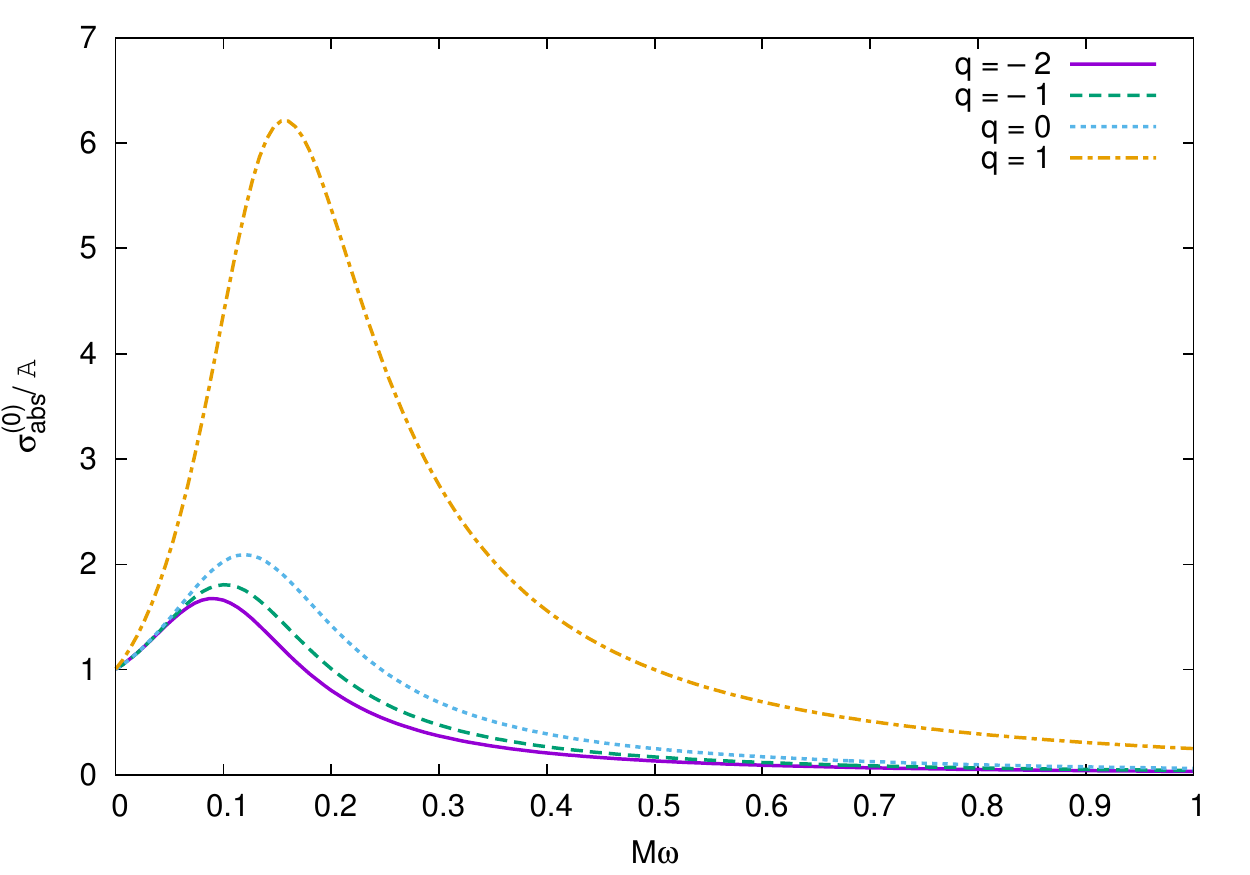}
 \caption{Partial absorption cross sections in units of the black hole mass squared for $l = 0,1$ (left) and in units of the black hole area for $l = 0$ (right).}
 \label{fig:pacs}
\end{figure}

Figure~\ref{fig:pacs} shows the partial absorption cross sections for black holes with tidal charge $q = -1,-2$ and also for the Schwarzschild and extreme Reissner--Nordström black holes. We compare the partial cross section for $l = 0$ with both the black hole mass squared (left graph) and the black hole area, $\mathcal{A} = 4\pi r_h^2$ (right graph). For fixed mass, the partial absorption cross section increases with the decrease of $q$. For black holes with fixed area, however, the partial absorption cross section rapidly increases with the increase of $q$. This is so because the black hole tends to shrink as $q$ becomes higher. Therefore, in the right panel of fig.~\ref{fig:pacs}, we are actually comparing black holes with different masses which are higher for higher $q$ in order to keep the event horizon size unaltered. Also from the right panel of fig.~\ref{fig:pacs} we can infer that $\sigma_\text{abs} \to \mathcal{A}$ when $M\omega \to 0$.\footnote{Although the right panel of fig.~\ref{fig:pacs} refers to the partial absorption cross section with $l = 0$, $\sigma_l \to 0$ for $l > 0$ when $M\omega \to 0$ and therefore the non-vanishing contribution to the total cross section in the zero-frequency limit comes only from the $l = 0$ mode.} This is in agreement with analytical results which predict that stationary black holes have the absorption cross section for low-frequency massless scalar field equal to their area~\cite{Higuchi2001cqg18_L139,Das1997prl78_417}. In the left graph of fig.~\ref{fig:pacs} we also show the partial absorption cross sections for $l = 1$ in units of the black hole mass squared. Again the cross section increases with the decrease of $q$, as expected from the effective potential behavior (see fig.~\ref{fig:V}). Similar results have been observed for higher values of $l$.

\begin{figure}[!htb]
 \centering
 \includegraphics[width=0.49\textwidth]{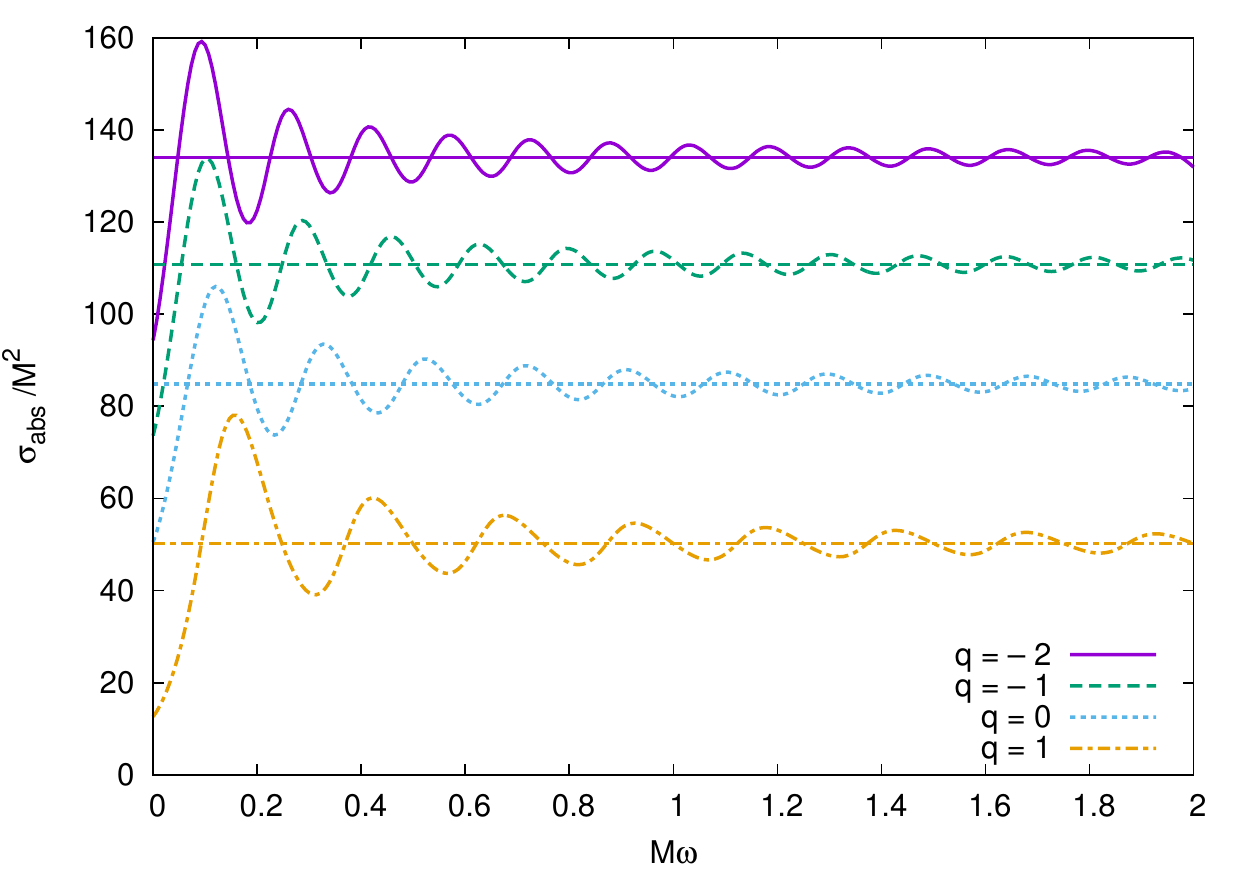}
 \includegraphics[width=0.49\textwidth]{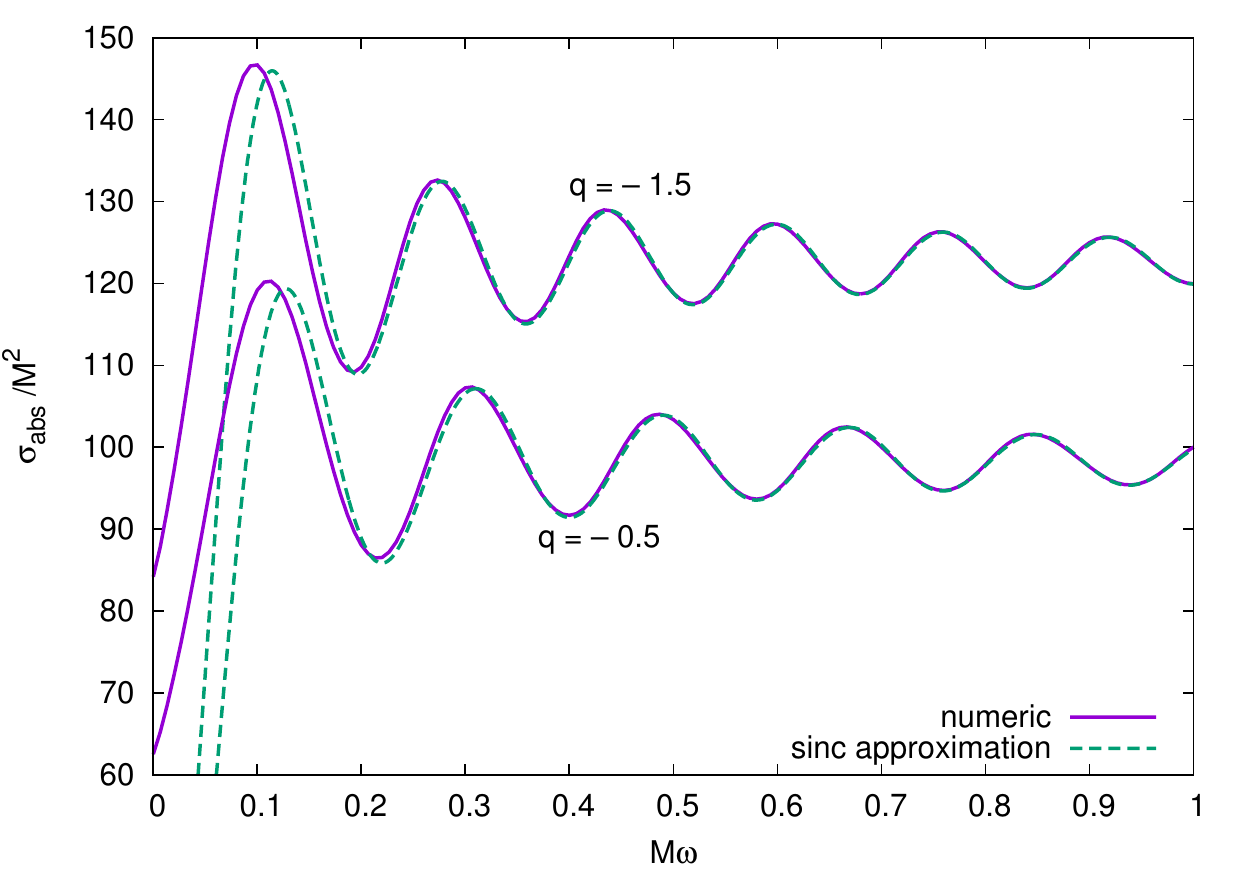}
 \caption{Total scalar absorption cross section of black holes with tidal charges $q = -2,-1$, Schwarzschild ($q = 0$) and extreme Reissner--Nordström ($q = 1$) black holes compared with the geometrical-optics limit (left) and for $q = -0.5,-1.5$ compared with the ``sinc approximation''~\cite{Toshmatov2016prd93_124017} (right).}
 \label{fig:tacs}
\end{figure}

In fig.~\ref{fig:tacs} we present the results for the total absorption cross sections. Left panel shows the comparison of the total absorption cross sections for black holes with tidal charge $q = -2,-1$ as well as for Schwarzschild and extreme Reissner--Nordström black holes considering the massless scalar field. In the right panel we compare the numeric results with the ``sinc approximation''~\cite{Decanini2011prd83_044032} for $q = -0.5, -1.5$ which were obtained and originally presented in ref.~\cite{Toshmatov2016prd93_124017} (see fig. 9 therein) and are reproduced here thanks to their authors. We see that the absorption cross section rapidly decreases with the increase of $q$, as has been already observed for the partial absorption in the left graph of fig.~\ref{fig:pacs}. Also, the absorption cross section oscillations become wider with the increase of $q$. The wave absorption cross section oscillates around the geometrical optics limit value (straight lines on the left panel) and excellently agree with the ``sinc approximation'' already for relatively low values of the frequency, $M\omega \sim 0.5$.

\begin{figure}[!htb]
\centering
\includegraphics[width=\textwidth]{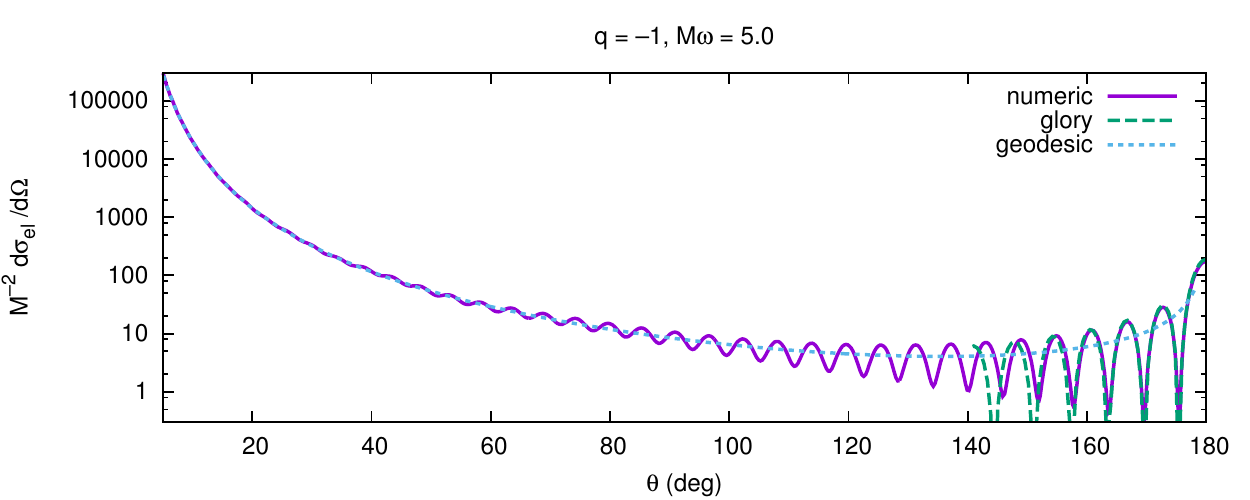}
\includegraphics[width=\textwidth]{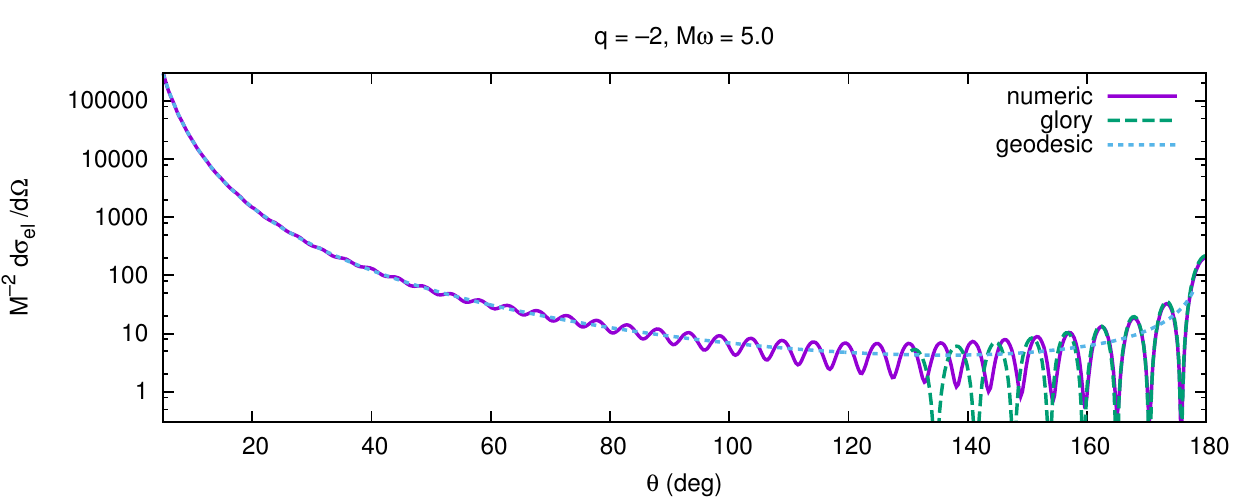}
\caption{Comparison of the scalar differential scattering cross section with geodesic and glory approximations for $q = -1$ (top) and $q = -2$ (bottom).}
\label{fig:comp-approx}
\end{figure}

Figure~\ref{fig:comp-approx} shows the comparison of the differential scattering cross sections of black holes with $q= -2,-1$ and $M\omega = 5.0$ obtained numerically, via geodesic approach, and via the glory approximation for massless scalar particles. In all cases we see that the glory approximation fits well with numeric results for $\theta \gtrsim 160^\circ$ while the classical result is already very close to the numeric results in the range $\theta \lesssim 20^\circ$.

\begin{figure}[!htb]
 \centering
 \includegraphics[width=\textwidth]{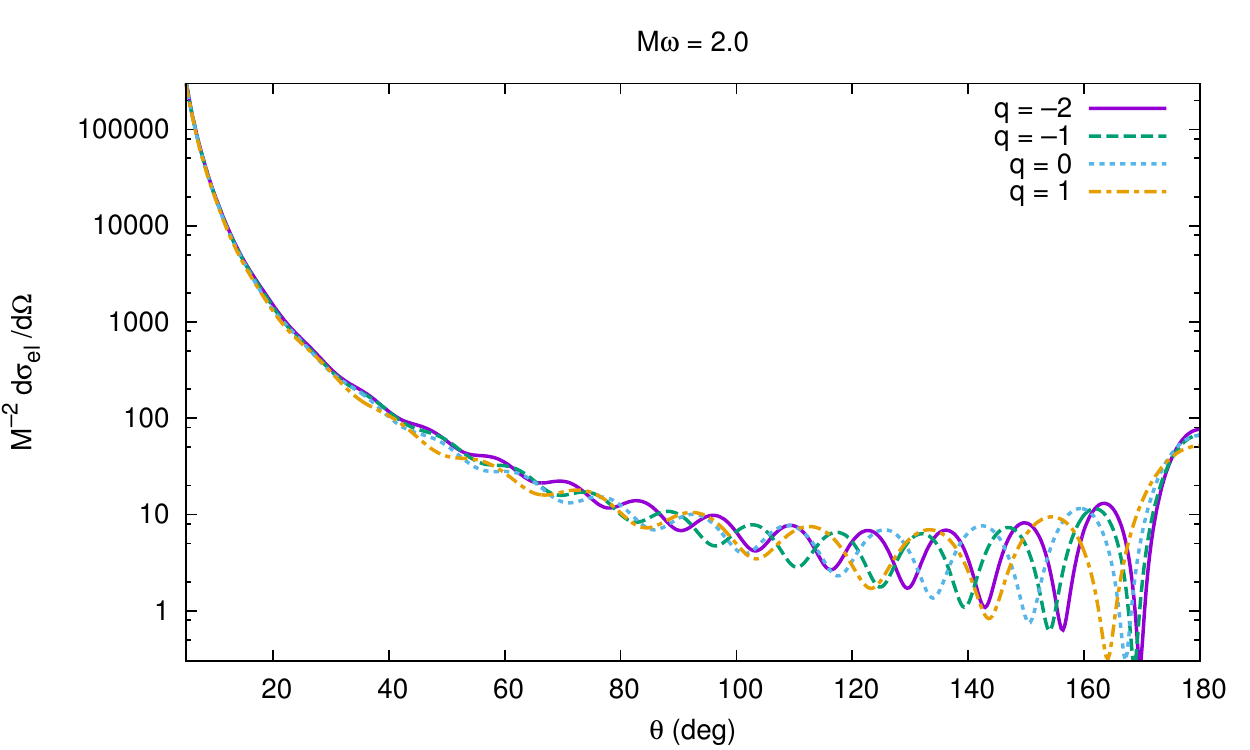}
 \caption{Scalar differential scattering cross sections of black holes with $q = -2,-1,0,1$. The cases $q=0,1$ correspond respectively to the Schwarzschild and the extreme Reissner--Nordström black holes.}
 \label{fig:dscs-q_comp}
\end{figure}

A comparison of the differential scattering cross section of black holes with tidal charge $q = -2,-1$, Schwarzschild ($q = 0$) and extreme Reissner--Nordström ($q = 1$) black holes for the massless scalar field with $M\omega = 2.0$ is presented in fig.~\ref{fig:dscs-q_comp}. We see that the tidal charge does not play an important role in the average of the scattered-flux intensities, but its change modifies the width of interference fringes of the differential scattering cross section. The decrease of $q$ implies in a decrease of the interference fringe widths. Near the forward direction, $\theta \lesssim 20^\circ$, the results tend to depend weakly on the value of $q$. This is expected since the metric term of the charge $q$ varies with $r^{-2}$ and tends to be small compared with the mass term in the far region where particles which suffer small deflections pass.

\section{Final remarks\label{sec:remarks}}

We have computed the absorption and scattering cross sections of black holes with tidal charges for the massless scalar field. Since the metric form of these black holes coincides with the metric form of Reissner--Nordström black holes, we have compared the main results with similar results for both Schwarzschild and extreme Reissner--Nordström black holes to have a clearer understanding of the effect of the tidal charge in the cross sections.

The tidal charge acts mainly in phenomena which take place in the region near the black hole. Black holes with more intense tidal charges, e.g. higher $-q$, tend to absorb more, as noticed by the increase of the partial and total absorption cross sections when presented in units of the black hole mass squared (cf. left graph of fig.~\ref{fig:pacs} and fig.~\ref{fig:tacs}). The same can be conclude by analyzing the comparison of the differential scattering cross section for different values of $q$ (cf. fig.~\ref{fig:dscs-q_comp}). In this case, we have shown that a change in $q$ has direct consequences on the interference fringe widths, which are more intense for high values of the scattering angle. In the near-forward direction, small $\theta$, the interference fringes wane, and the differential scattering cross sections tend to be the same, not depending on the value of $q$. Similar consequences of the change of $q$ were noticed in the case of Reissner--Nordström black holes~\cite{Jung2005npb717_272,Crispino2009prd79_064022}.

All approximations regarding to the cross sections of the massless scalar field apply well in the case of black holes with tidal charge. We have shown that such black holes obey the universality of the low-frequency absorption -- which says that the absorption cross section for the massles scalar field tends to the black hole area in the low-energy limit if the black hole is stationary~\cite{Higuchi2001cqg18_L139} -- by expressing numeric results in terms of the black hole area (cf. right graph of fig.~\ref{fig:pacs}). We also showed that the total absorption cross sections tend to oscillate around the corresponding capture cross sections in the geometrical-optics limit (cf. left graph of fig.~\ref{fig:tacs}) and excellently agree with the ``sinc approximation''~\cite{Toshmatov2016prd93_124017} (see right panel of fig.~\ref{fig:tacs}) which is valid in the high-frequency limit~\cite{Decanini2011prd83_044032}. In the case of the differential scattering cross sections, the analytical results, geodesic and glory approximations, were shown to be in good agreement with the numeric results in their respective regime of validity, low scattering angles for geodesics and large angles for the glory approximation, even though not very high frequencies have been considered.

\section*{Acknowledgments}

The author would like to thank Conselho Nacional de Desenvolvimento Científico e Tecnológico
(CNPq) for partial financial support via the grant 310911/2015-0, to B. Toshmatov, Z. Stuchlík, J. Schee, and B. Ahmedov who allowed the reproduction of their results for the ``sinc approximation'' and to B. Toshmatov for kindly sending the files necessary to generate the curves of this approximation which are shown here.
%

\begin{thebibliography}{61}
\providecommand{\natexlab}[1]{#1}
\providecommand{\url}[1]{\texttt{#1}}
\expandafter\ifx\csname urlstyle\endcsname\relax
  \providecommand{\doi}[1]{doi: #1}\else
  \providecommand{\doi}{doi: \begingroup \urlstyle{rm}\Url}\fi

\bibitem[Arkani-Hamed et~al.(1998)Arkani-Hamed, Dimopoulos, and
  Dvali]{Arkani1998plb429_263}
Nima Arkani-Hamed, Savas Dimopoulos, and Gia Dvali.
\newblock {The hierarchy problem and new dimensions at a millimeter}.
\newblock \emph{Phys. Lett. B}, 429:\penalty0 263--272, 1998.
\newblock \doi{10.1016/S0370-2693(98)00466-3}.

\bibitem[Antoniadis et~al.(1998)Antoniadis, Arkani-Hamed, Dimopoulos, and
  Dvali]{Antoniadis1998plb436_257}
Ignatios Antoniadis, Nima Arkani-Hamed, Savas Dimopoulos, and Gia Dvali.
\newblock {New dimensions at a millimeter to a fermi and superstrings at a
  TeV}.
\newblock \emph{Phys. Lett. B}, 436:\penalty0 257--263, 1998.
\newblock \doi{10.1016/S0370-2693(98)00860-0}.

\bibitem[Randall and Sundrum(1999{\natexlab{a}})]{Randall1999prl83_3370}
Lisa Randall and Raman Sundrum.
\newblock {Large Mass Hierarchy from a Small Extra Dimension}.
\newblock \emph{Phys. Rev. Lett.}, 83:\penalty0 3370--3373, 1999{\natexlab{a}}.
\newblock \doi{10.1103/PhysRevLett.83.3370}.

\bibitem[Randall and Sundrum(1999{\natexlab{b}})]{Randall1999prl83_4690}
Lisa Randall and Raman Sundrum.
\newblock {An Alternative to Compactification}.
\newblock \emph{Phys. Rev. Lett.}, 83:\penalty0 4690--4693, 1999{\natexlab{b}}.
\newblock \doi{10.1103/PhysRevLett.83.4690}.

\bibitem[Park(2012)]{Park2012ppnp67_617}
Seong~Chan Park.
\newblock {Black holes and the LHC: A review}.
\newblock \emph{Prog. Part. Nucl. Phys.}, 67:\penalty0 617--650, 2012.
\newblock \doi{10.1016/j.ppnp.2012.03.004}.

\bibitem[Collaboration(2016{\natexlab{a}})]{ATLAS2016jhep03_041}
ATLAS Collaboration.
\newblock {Search for new phenomena with photon+jet events in proton-proton
  collisions at $ \sqrt{s}=13 $ TeV with the ATLAS detector}.
\newblock \emph{JHEP}, 03:\penalty0 041, 2016{\natexlab{a}}.
\newblock \doi{10.1007/JHEP03(2016)041}.

\bibitem[Collaboration(2016{\natexlab{b}})]{ATLAS2016plb754_302}
ATLAS Collaboration.
\newblock {Search for new phenomena in dijet mass and angular distributions
  from $pp$ collisions at $\sqrt{s}=$ 13 TeV with the ATLAS detector}.
\newblock \emph{Phys. Lett. B}, 754:\penalty0 302--322, 2016{\natexlab{b}}.
\newblock \doi{10.1016/j.physletb.2016.01.032}.

\bibitem[Collaboration(2017{\natexlab{a}})]{CMS2017plb774_279}
CMS Collaboration.
\newblock {Search for black holes and other new phenomena in high-multiplicity
  final states in proton–proton collisions at $ \sqrt{s}=$13 TeV}.
\newblock \emph{Phys. Lett. B}, 774:\penalty0 279--307, 2017{\natexlab{a}}.
\newblock \doi{10.1016/j.physletb.2017.09.053}.

\bibitem[Collaboration(2017{\natexlab{b}})]{ATLAS2017prd96_052004}
ATLAS Collaboration.
\newblock {Search for new phenomena in dijet events using 37 fb$^{-1}$ of $pp$
  collision data collected at $\sqrt{s}=$13 TeV with the ATLAS detector}.
\newblock \emph{Phys. Rev. D}, 96\penalty0 (5):\penalty0 052004,
  2017{\natexlab{b}}.
\newblock \doi{10.1103/PhysRevD.96.052004}.

\bibitem[Collaboration(2018)]{ATLAS2018epjc78_102}
ATLAS Collaboration.
\newblock {Search for new phenomena in high-mass final states with a photon and
  a jet from $pp$ collisions at $\sqrt{s}$ = 13 TeV with the ATLAS detector}.
\newblock \emph{Eur. Phys. J. C}, 78\penalty0 (2):\penalty0 102, 2018.
\newblock \doi{10.1140/epjc/s10052-018-5553-2}.

\bibitem[Hawking(1975)]{Hawking1975cmp43_199}
S.~W. Hawking.
\newblock {Particle creation by black holes}.
\newblock \emph{Commun. Math. Phys.}, 43:\penalty0 199--220, 1975.
\newblock \doi{10.1007/BF02345020}.
\newblock [,167(1975)].

\bibitem[Emparan et~al.(2000)Emparan, Horowitz, and
  Myers]{Emparan2000prl85_499}
Roberto Emparan, Gary~T. Horowitz, and Robert~C. Myers.
\newblock {Black Holes Radiate Mainly on the Brane}.
\newblock \emph{Phys. Rev. Lett.}, 85:\penalty0 499--502, 2000.
\newblock \doi{10.1103/PhysRevLett.85.499}.

\bibitem[Kanti and March-Russell(2002)]{Kanti2002prd66_024023}
Panagiota Kanti and John March-Russell.
\newblock {Calculable corrections to brane black hole decay: The scalar case}.
\newblock \emph{Phys. Rev. D}, 66:\penalty0 024023, 2002.
\newblock \doi{10.1103/PhysRevD.66.024023}.

\bibitem[Kanti and March-Russell(2003)]{Kanti2003prd67_104019}
Panagiota Kanti and John March-Russell.
\newblock {Calculable corrections to brane black hole decay. II. Greybody
  factors for spin 1/2 and 1}.
\newblock \emph{Phys. Rev. D}, 67:\penalty0 104019, 2003.
\newblock \doi{10.1103/PhysRevD.67.104019}.

\bibitem[Harris and Kanti(2003)]{Harris2003jhep10_014}
Christopher~M. Harris and Panagiota Kanti.
\newblock {Hawking radiation from a $(4+n)$-dimensional black hole: exact
  results for the Schwarzschild phase}.
\newblock \emph{JHEP}, 10:\penalty0 014, 2003.
\newblock \doi{10.1088/1126-6708/2003/10/014}.

\bibitem[Jung et~al.(2004)Jung, Kim, and Park]{Jung2004jhep09_005}
Eylee Jung, SungHoon Kim, and D.~K. Park.
\newblock {Low-energy absorption cross section for massive scalar and Dirac
  fermion by (4+n)-dimensional Schwarzschild black hole}.
\newblock \emph{JHEP}, 09:\penalty0 005, 2004.
\newblock \doi{10.1088/1126-6708/2004/09/005}.

\bibitem[Kanti et~al.(2005)Kanti, Grain, and Barrau]{Kanti2005prd71_104002}
P.~Kanti, J.~Grain, and A.~Barrau.
\newblock {Bulk and brane decay of a $(4+n)$-dimensional Schwarzschild–de
  Sitter black hole: Scalar radiation}.
\newblock \emph{Phys. Rev. D}, 71:\penalty0 104002, 2005.
\newblock \doi{10.1103/PhysRevD.71.104002}.

\bibitem[Ida et~al.(2003)Ida, Oda, and Park]{Ida2003prd67_064025}
Daisuke Ida, Kin-ya Oda, and Seong~Chan Park.
\newblock {Rotating black holes at future colliders: Greybody factors for brane
  fields}.
\newblock \emph{Phys. Rev. D}, 67:\penalty0 064025, 2003.
\newblock \doi{10.1103/PhysRevD.67.064025, 10.1103/PhysRevD.69.049901}.
\newblock [Erratum: Phys. Rev.D69,049901(2004)].

\bibitem[Ida et~al.(2005)Ida, Oda, and Park]{Ida2005prd71_124039}
Daisuke Ida, Kin-ya Oda, and Seong~Chan Park.
\newblock {Rotating black holes at future colliders. II. Anisotropic scalar
  field emission}.
\newblock \emph{Phys. Rev. D}, 71:\penalty0 124039, 2005.
\newblock \doi{10.1103/PhysRevD.71.124039}.

\bibitem[Ida et~al.(2006)Ida, Oda, and Park]{Ida2006prd73_124022}
Daisuke Ida, Kin-ya Oda, and Seong~Chan Park.
\newblock {Rotating black holes at future colliders. III. Determination of
  black hole evolution}.
\newblock \emph{Phys. Rev.}, D73:\penalty0 124022, 2006.
\newblock \doi{10.1103/PhysRevD.73.124022}.

\bibitem[Harris and Kanti(2006)]{Harris2005plb633_106}
C.~M. Harris and P.~Kanti.
\newblock {Hawking radiation from a $(4+n)$-dimensional rotating black hole on
  the brane}.
\newblock \emph{Phys. Lett. B}, 633:\penalty0 106--110, 2006.
\newblock \doi{10.1016/j.physletb.2005.10.025}.

\bibitem[Duffy et~al.(2005)Duffy, Harris, Kanti, and
  Winstanley]{Duffy2005jhep09_049}
Gavin Duffy, Christopher~M. Harris, Panagiota Kanti, and Elizabeth Winstanley.
\newblock {Brane decay of a (4+n)-dimensional rotating black hole: Spin-0
  particles}.
\newblock \emph{JHEP}, 09:\penalty0 049, 2005.
\newblock \doi{10.1088/1126-6708/2005/09/049}.

\bibitem[Casals et~al.(2006)Casals, Kanti, and
  Winstanley]{Casals2006jhep02_051}
Marc Casals, Panagiota Kanti, and Elizabeth Winstanley.
\newblock {Brane decay of a $(4+n)$-dimensional rotating black hole II: spin-1
  particles}.
\newblock \emph{JHEP}, 02:\penalty0 051, 2006.
\newblock \doi{10.1088/1126-6708/2006/02/051}.

\bibitem[Casals et~al.(2007)Casals, Dolan, Kanti, and
  Winstanley]{Casals2006jhep03_019}
Marc Casals, Sam Dolan, Panagiota Kanti, and Elizabeth Winstanley.
\newblock {Brane decay of a $(4+n)$-dimensional rotating black hole. III:
  spin-1/2 particles}.
\newblock \emph{JHEP}, 03:\penalty0 019, 2007.
\newblock \doi{10.1088/1126-6708/2007/03/019}.

\bibitem[Creek et~al.(2007)Creek, Efthimiou, Kanti, and
  Tamvakis]{Creek2007prd75_084043}
S.~Creek, O.~Efthimiou, P.~Kanti, and K.~Tamvakis.
\newblock {Greybody factors for brane scalar fields in a rotating black hole
  background}.
\newblock \emph{Phys. Rev. D}, 75:\penalty0 084043, 2007.
\newblock \doi{10.1103/PhysRevD.75.084043}.

\bibitem[Kanti et~al.(2014)Kanti, Pappas, and Pappas]{Kanti2014prd90_124077}
P.~Kanti, T.~Pappas, and N.~Pappas.
\newblock {Greybody factors for scalar fields emitted by a higher-dimensional
  Schwarzschild–de Sitter black hole}.
\newblock \emph{Phys. Rev. D}, 90\penalty0 (12):\penalty0 124077, 2014.
\newblock \doi{10.1103/PhysRevD.90.124077}.

\bibitem[Majumdar and Mukherjee(2005)]{Majumdar2005ijmpd14_1095}
A.~S. Majumdar and N.~Mukherjee.
\newblock {Braneworld black holes in cosmology and astrophysics}.
\newblock \emph{Int. J. Mod. Phys. D}, 14:\penalty0 1095, 2005.
\newblock \doi{10.1142/S0218271805006948}.

\bibitem[Bronnikov et~al.(2003)Bronnikov, Melnikov, and
  Dehnen]{Bronnikov2003prd68_024025}
K.~A. Bronnikov, V.~N. Melnikov, and Heinz Dehnen.
\newblock {General class of brane-world black holes}.
\newblock \emph{Phys. Rev. D}, 68:\penalty0 024025, 2003.
\newblock \doi{10.1103/PhysRevD.68.024025}.

\bibitem[Marinho and de~Oliveira(2016)]{Marinho2016arxiv1612_05604}
C\'assio I.~S. Marinho and Ednilton~S. de~Oliveira.
\newblock {Scattering of massless scalar waves from Schwarzschild-Tangherlini
  black holes on the brane}.
\newblock \emph{arXiv}, page 1612.05604, 2016.

\bibitem[de~Oliveira(2018)]{deOliveira2017cqg35_065007}
Ednilton~S. de~Oliveira.
\newblock {Scalar scattering from charged black holes on the brane}.
\newblock \emph{Class. Quant. Grav.}, 35\penalty0 (6):\penalty0 065007, 2018.
\newblock \doi{10.1088/1361-6382/aaa5c2}.

\bibitem[Jung and Park(2005)]{Jung2005npb717_272}
Eylee Jung and D.~K. Park.
\newblock {Absorption and emission spectra of an higher-dimensional
  Reissner-Nordström black hole}.
\newblock \emph{Nucl. Phys. B}, 717:\penalty0 272--303, 2005.
\newblock \doi{10.1016/j.nuclphysb.2005.03.037}.

\bibitem[Toshmatov et~al.(2016)Toshmatov, Stuchlík, Schee, and
  Ahmedov]{Toshmatov2016prd93_124017}
Bobir Toshmatov, Zdeněk Stuchlík, Jan Schee, and Bobomurat Ahmedov.
\newblock {Quasinormal frequencies of black hole in the braneworld}.
\newblock \emph{Phys. Rev. D}, 93\penalty0 (12):\penalty0 124017, 2016.
\newblock \doi{10.1103/PhysRevD.93.124017}.

\bibitem[Schee and Stuchlík(2009)]{Schee2008ijmpd18_983}
Jan Schee and Zdeněk Stuchlík.
\newblock {Optical phenomena in the field of braneworld Kerr black holes}.
\newblock \emph{Int. J. Mod. Phys. D}, 18:\penalty0 983--1024, 2009.
\newblock \doi{10.1142/S0218271809014881}.

\bibitem[Amarilla and Eiroa(2012)]{Amarilla2011pprd85_064019}
Leonardo Amarilla and Ernesto~F. Eiroa.
\newblock {Shadow of a rotating braneworld black hole}.
\newblock \emph{Phys. Rev. D}, 85:\penalty0 064019, 2012.
\newblock \doi{10.1103/PhysRevD.85.064019}.

\bibitem[Abdujabbarov et~al.(2017)Abdujabbarov, Ahmedov, Dadhich, and
  Atamurotov]{Abdujabbarov2017prd96_084017}
Ahmadjon Abdujabbarov, Bobomurat Ahmedov, Naresh Dadhich, and Farruh
  Atamurotov.
\newblock {Optical properties of a braneworld black hole: Gravitational lensing
  and retrolensing}.
\newblock \emph{Phys. Rev. D}, 96\penalty0 (8):\penalty0 084017, 2017.
\newblock \doi{10.1103/PhysRevD.96.084017}.

\bibitem[Eiroa and Sendra(2018)]{Eiroa2017epjc78_91}
Ernesto~F. Eiroa and Carlos~M. Sendra.
\newblock {Shadow cast by rotating braneworld black holes with a cosmological
  constant}.
\newblock \emph{Eur. Phys. J. C}, 78\penalty0 (2):\penalty0 91, 2018.
\newblock \doi{10.1140/epjc/s10052-018-5586-6}.

\bibitem[Kar and Sinha(2003)]{Kar2003grg35_1775}
Sayan Kar and Manodeep Sinha.
\newblock {Bending of light and gravitational signals in certain on-brane and
  bulk geometries}.
\newblock \emph{Gen. Rel. Grav.}, 35:\penalty0 1775--1784, 2003.
\newblock \doi{10.1023/A:1026057929329}.

\bibitem[Kanti and Konoplya(2006)]{Kanti2006prd73_044002}
P.~Kanti and R.~A. Konoplya.
\newblock {Quasinormal modes of brane-localized standard model fields}.
\newblock \emph{Phys. Rev. D}, 73:\penalty0 044002, 2006.
\newblock \doi{10.1103/PhysRevD.73.044002}.

\bibitem[Kanti et~al.(2006)Kanti, Konoplya, and
  Zhidenko]{Kanti2006prd74_064008}
P.~Kanti, R.~A. Konoplya, and A.~Zhidenko.
\newblock {Quasinormal modes of brane-localized standard model fields. II. Kerr
  black holes}.
\newblock \emph{Phys. Rev. D}, 74:\penalty0 064008, 2006.
\newblock \doi{10.1103/PhysRevD.74.064008}.

\bibitem[Molina et~al.(2016)Molina, Pavan, and
  Medina~Torrejón]{Molina2016prd93_124068}
C.~Molina, A.~B. Pavan, and T.~E. Medina~Torrejón.
\newblock {Electromagnetic perturbations in new brane world scenarios}.
\newblock \emph{Phys. Rev. D}, 93\penalty0 (12):\penalty0 124068, 2016.
\newblock \doi{10.1103/PhysRevD.93.124068}.

\bibitem[Dadhich et~al.(2000)Dadhich, Maartens, Papadopoulos, and
  Rezania]{Dadhich2000plb487_1}
Naresh Dadhich, Roy Maartens, Philippos Papadopoulos, and Vahid Rezania.
\newblock {Black holes on the brane}.
\newblock \emph{Phys. Lett. B}, 487:\penalty0 1--6, 2000.
\newblock \doi{10.1016/S0370-2693(00)00798-X}.

\bibitem[Chandrasekhar(1983)]{Chandra1983}
Subrahmanyan Chandrasekhar.
\newblock \emph{The Mathematical Theory of Black Holes}.
\newblock Clarendon Press, Oxford, 1983.
\newblock ISBN 0-19-851291-0.

\bibitem[Abramowitz and Stegun(1965)]{Abramowitz_etal1964}
M.~Abramowitz and I.~A. Stegun.
\newblock \emph{Handbook of Mathematical Functions}.
\newblock Dover Publications, New York, 1965.

\bibitem[Crispino et~al.(2009)Crispino, Dolan, and
  Oliveira]{Crispino2009prd79_064022}
Luís C.~B. Crispino, Sam~R. Dolan, and Ednilton~S. Oliveira.
\newblock {Scattering of massless scalar waves by Reissner-Nordstr\"om black
  holes}.
\newblock \emph{Phys. Rev. D}, 79:\penalty0 064022, 2009.
\newblock \doi{10.1103/PhysRevD.79.064022}.

\bibitem[DeWitt-Morette and Nelson(1984)]{DeWitt-Morette1984prd29_1663}
Cécile DeWitt-Morette and Bruce~L. Nelson.
\newblock {Glories---and other degenerate points of the action}.
\newblock \emph{Phys. Rev. D}, 29:\penalty0 1663--1668, 1984.
\newblock \doi{10.1103/PhysRevD.29.1663}.

\bibitem[Futterman et~al.(1988)Futterman, Handler, and
  Matzner]{Futterman_etal1988}
J.~A.~H. Futterman, F.~A. Handler, and R.~A. Matzner.
\newblock \emph{Scattering from Black Holes}.
\newblock Cambridge University Press, Cambridge, 1988.
\newblock ISBN 0-521-32986-8.

\bibitem[Gradshteyn and Ryzhik(2015)]{Gradshteyn_etal2000}
I.~S. Gradshteyn and I.~M. Ryzhik.
\newblock \emph{Table of Integrals, Series, and Products}.
\newblock Academic Press, San Diego, 8 th edition, 2015.
\newblock ISBN 978-0-12-384933-5.

\bibitem[Crispino et~al.(2014)Crispino, Dolan, Higuchi, and
  de~Oliveira]{Crispino2014prd90_064027}
Luís C.~B. Crispino, Sam~R. Dolan, Atsushi Higuchi, and Ednilton~S.
  de~Oliveira.
\newblock {Inferring black hole charge from backscattered electromagnetic
  radiation}.
\newblock \emph{Phys. Rev. D}, 90\penalty0 (6):\penalty0 064027, 2014.
\newblock \doi{10.1103/PhysRevD.90.064027}.

\bibitem[Crispino et~al.(2015)Crispino, Dolan, Higuchi, and
  de~Oliveira]{Crispino2015prd92_084056}
Luís C.~B. Crispino, Sam~R. Dolan, Atsushi Higuchi, and Ednilton~S.
  de~Oliveira.
\newblock {Scattering from charged black holes and supergravity}.
\newblock \emph{Phys. Rev. D}, 92\penalty0 (8):\penalty0 084056, 2015.
\newblock \doi{10.1103/PhysRevD.92.084056}.

\bibitem[Sanchez(1976)]{Sanchez1976jmp17_688}
Norma~G. Sanchez.
\newblock {Scattering of scalar waves from a Schwarzschild black hole}.
\newblock \emph{J. Math. Phys.}, 17\penalty0 (5):\penalty0 688, 1976.
\newblock \doi{10.1063/1.522949}.

\bibitem[Sánchez(1977)]{Sanchez1976prd16_937}
Norma Sánchez.
\newblock {Wave scattering theory and the absorption problem for a black hole}.
\newblock \emph{Phys. Rev. D}, 16:\penalty0 937--945, 1977.
\newblock \doi{10.1103/PhysRevD.16.937}.

\bibitem[Sanchez(1978)]{Sanchez1978prd18_1030}
Norma Sanchez.
\newblock {Absorption and emission spectra of a Schwarzschild black hole}.
\newblock \emph{Phys. Rev. D}, 18:\penalty0 1030, 1978.
\newblock \doi{10.1103/PhysRevD.18.1030}.

\bibitem[Crispino et~al.(2007)Crispino, Oliveira, and
  Matsas]{Crispino2007prd76_107502}
Luís C.~B. Crispino, Ednilton~S. Oliveira, and George E.~A. Matsas.
\newblock {Absorption cross section of canonical acoustic holes}.
\newblock \emph{Phys. Rev. D}, 76:\penalty0 107502, 2007.
\newblock \doi{10.1103/PhysRevD.76.107502}.

\bibitem[Macedo and Crispino(2014)]{Macedo2014prd90_064001}
Caio F.~B. Macedo and Luís C.~B. Crispino.
\newblock {Absorption of planar massless scalar waves by Bardeen regular black
  holes}.
\newblock \emph{Phys. Rev. D}, 90\penalty0 (6):\penalty0 064001, 2014.
\newblock \doi{10.1103/PhysRevD.90.064001}.

\bibitem[Sánchez(1978)]{Sanchez1978prd18_1798}
Norma Sánchez.
\newblock {Elastic scattering of waves by a black hole}.
\newblock \emph{Phys. Rev. D}, 18:\penalty0 1798, 1978.
\newblock \doi{10.1103/PhysRevD.18.1798}.

\bibitem[Dolan et~al.(2009)Dolan, Oliveira, and
  Crispino]{Dolan2009prd79_064014}
Sam~R. Dolan, Ednilton~S. Oliveira, and Luís C.~B. Crispino.
\newblock {Scattering of sound waves by a canonical acoustic hole}.
\newblock \emph{Phys. Rev. D}, 79:\penalty0 064014, 2009.
\newblock \doi{10.1103/PhysRevD.79.064014}.

\bibitem[Macedo et~al.(2015)Macedo, de~Oliveira, and
  Crispino]{Macedo2015prd91_024012}
Caio F.~B. Macedo, Ednilton~S. de~Oliveira, and Luís C.~B. Crispino.
\newblock {Scattering by regular black holes: Planar massless scalar waves
  impinging upon a Bardeen black hole}.
\newblock \emph{Phys. Rev. D}, 92\penalty0 (2):\penalty0 024012, 2015.
\newblock \doi{10.1103/PhysRevD.92.024012}.

\bibitem[Yennie et~al.(1954)Yennie, Ravenhall, and Wilson]{Yennie1954pr85_500}
D.~R. Yennie, D.~G. Ravenhall, and R.~N. Wilson.
\newblock {Phase-Shift Calculation of High-Energy Electron Scattering}.
\newblock \emph{Phys. Rev.}, 95:\penalty0 500--512, 1954.
\newblock \doi{10.1103/PhysRev.95.500}.

\bibitem[Higuchi(2001)]{Higuchi2001cqg18_L139}
Atsushi Higuchi.
\newblock {Low-frequency scalar absorption cross sections for stationary black
  holes}.
\newblock \emph{Class. Quant. Grav.}, 18:\penalty0 L139, 2001.
\newblock \doi{10.1088/0264-9381/18/20/102}.
\newblock [Addendum: Class. Quant. Grav.19,599(2002)].

\bibitem[Das et~al.(1997)Das, Gibbons, and Mathur]{Das1997prl78_417}
Sumit~R. Das, Gary Gibbons, and Samir~D. Mathur.
\newblock {Universality of Low Energy Absorption Cross Sections for Black
  Holes}.
\newblock \emph{Phys. Rev. Lett.}, 78:\penalty0 417--419, 1997.
\newblock \doi{10.1103/PhysRevLett.78.417}.

\bibitem[Décanini et~al.(2011)Décanini, Esposito-Farèse, and
  Folacci]{Decanini2011prd83_044032}
Yves Décanini, Gilles Esposito-Farèse, and Antoine Folacci.
\newblock {Universality of high-energy absorption cross sections for black
  holes}.
\newblock \emph{Phys. Rev. D}, 83:\penalty0 044032, 2011.
\newblock \doi{10.1103/PhysRevD.83.044032}.

\end{thebibliography}

\end{document}